\documentclass[letterpaper, 10 pt, conference]{ieeeconf}  

\IEEEoverridecommandlockouts                              
\usepackage[dvipdfmx]{graphicx}
\usepackage{subfigure}
\usepackage{amsmath}
\usepackage[varg]{txfonts}
\usepackage{latexsym}
\newtheorem{theorem}{Theorem}
\newtheorem{lemma}{Lemma}

\newtheorem{proposition}{Proposition}
\title{\LARGE \bf
Game Theoretic Approach to the Stabilization of Heterogeneous Multiagent Systems Using Subsidy
}

\author{Takuya Morimoto, Takafumi Kanazawa, and Toshimitsu Ushio
\thanks{T.\ Morimoto, T.\ Kanazawa, and T.\ Ushio are with Graduate School of Engineering Science, Osaka University, 1-3 Machikaneyama, Toyonaka, Osaka, 560-8531, Japan.}
\thanks{{\tt\small morimoto@hopf.sys.es.osaka-u.ac.jp}}
\thanks{{\tt\small kanazawa@sys.es.osaka-u.ac.jp}}
\thanks{{\tt\small ushio@sys.es.osaka-u.ac.jp}}%
}

\begin{document}

\maketitle
\thispagestyle{empty}
\pagestyle{empty}

\begin{abstract}

We consider a multiagent system consisting of selfish and heterogeneous agents. Its behavior is modeled by multipopulation replicator dynamics, where payoff functions of populations are different from each other. In general, there exist several equilibrium points in the replicator dynamics. In order to stabilize a desirable equilibrium point, we introduce a controller called a ``government'' which controls the behaviors of agents by offering them subsidies. In previous work, it is assumed that the government determines the subsidies based on the populations the agents belong to. In general, however, the government cannot identify the members of each population. In this paper, we assume that the government observes the action of each agent and determines the subsidies based on the observed action profile.  Then, we model the controlled behaviors of the agents using replicator dynamics with feedback. We derive a stabilization condition of the target equilibrium point in the replicator dynamics.

\end{abstract}

\section{Introduction}
Multiagent systems consist of a large number of agents interacting with each other. They offer an innovative way to manage large-scale, dynamic, and heterogeneous computing systems\ {\cite{key1}}. The internet and multi-database systems are examples of such systems\ {\cite{key2}}. The agents act selfishly so as to improve their own payoffs, which may be mutually different. In this paper, such a multiagent system is said to be heterogeneous. The agents are grouped into several populations and their payoff functions are same if they belong to the same population. Figure 1 illustrates the heterogeneous multiagent system. We model the systems by multipopulation replicator dynamics\ {\cite{key3}}. 
{\par}In general, there exist several equilibrium points in the replicator dynamics. However, the desirable equilibrium point is not always stable. For example, consider a situation where a multiagent system consists of a single population and two agents are repeatedly selected at random from the population to play a 2-player Prisoner's Dilemma game. All agents select ``defect'' or ``cooperate'', and change their actions based on their earned payoffs. In this case, there exist two equilibrium points: all agents select ``defect'', or all agents select ``cooperate''. The former is stable but undesirable, while the latter is unstable but desirable, since the latter earns a larger payoff than the former. Such a phenomenon is called a social dilemma and many approaches have been taken to resolve the dilemma\ {\cite{key4}}. Similar phenomena occur in the heterogeneous multiagent system. Thus, stabilization of an unstable but desirable equilibrium point in multiagent systems is an important issue. 
{\par}An approach to the stabilization problem is to improve agent's learning algorithm. Stimpson and Goodrich provide a satisficing algorithm which reaches a Pareto efficient solution in self-play, and avoids exploitation by selfish agents\ {\cite{key5}}. Moreover, the mechanism design is an effective method of solving the social dilemma\ {\cite{key6}}. The mechanism design is a theory to design the rules to achieve the desirable state in the case that the agents act selfishly\ {\cite{mecha}}.
{\par}In this paper, we introduce a controller called a ``government'' that stabilizes a desirable equilibrium point of the multiagent systems by offering the subsidies to agents. Kanazawa et al.\ consider a desirable equilibrium point as a target equilibrium point, and derive a condition of the amount of the subsidy for the target equilibrium point to be asymptotically stable in a homogeneous case\ {\cite{key7}}, and Ichiba et al.\ derive a condition in a heterogeneous case\ {\cite{key8}}. In the previous work\ {\cite{key8}}, it is assumed that the government knows which population each agent belongs to and determines the subsidies for agents based on both their populations and their actions in order to stabilize a target equilibrium point. However, the government cannot always identify the members of each population. Thus, we consider that the government determines the subsidies based on the observation of agents' actions. A stabilization condition for a target equilibrium point based on such a control action has been shown in 2-population and 2-action replicator dynamics\ {\cite{key9}}.
{\par}In this paper, we extend the condition to $m$-population and $n$-action replicator dynamics ($m\ {\geq}\ 2, n\ {\geq}\ 2)$. Figure 2 illustrates how the government controls agents. We assume that all agents have the same set of actions. Agents who belong to the same population have the same payoff functions depending on the populations. We assume that the government cannot identify the members of each population but can observe agents' actions (we call it an output). The government chooses a target output and offers the subsidies which depend on the output to achieve the target output. As a result, we consider a stabilization problem of a target equilibrium point which corresponds to the target output. Since the feedback action of the government is based on the output, we derive replicator dynamics based on the feedback. Then, we propose a control law for determining the subsidy for each agent, and show a sufficient condition for the global stabilization of the target output.   
{\par}This paper is organized as follows. In Section 2, we review the multipopulation replicator dynamics, which describes the behaviors of heterogeneous multiagent systems. In Section 3, we introduce a government that controls multiagent systems based on the observation of agents' actions, and derive a model to represent the behaviors of the controlled systems. In Section 4, we derive a stabilization condition of the target equilibrium point and show a numerical example.
\begin{figure}[t]
\begin{center}
\includegraphics[width=85mm,clip]{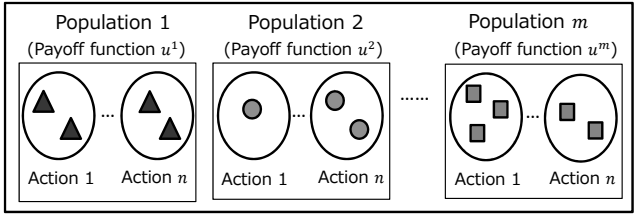}
\end{center}
\caption{Illustration of the heterogeneous multiagent system.}
\label{fig:one}
\end{figure}

\section{Multipopulation replicator dynamics}
Let us review a game theoretic model of the interaction among agents in multiple populations. We consider multiple populations that consist of a large number of agents. We assume that agents who belong to the same population have the same sense of values or the same criterion, while agents who belong to different populations have different ones. Since the sense of values or the criterion of each agent is represented as a payoff function, agents who belong to the same population have the same payoff function depending on the population. We also assume that, in a sufficiently long period of time, agents' senses of values or criteria do not change. All agents do not move among the populations, that is, the number of agents in each population is constant. We assume that all agents have the same set of actions. Two agents are repeatedly selected at random from all populations to play a 2-player game. They may belong to the same population. Agents change their actions depending on the payoffs earned in the game. Agents do not know which population their opponents belong to. In this paper, we use the following notations:
\noindent
\begin{itemize}
\item $P=\{1,\ldots,m\}$: the set of populations;
\item $S=\{1,\ldots,n\}$: the set of actions;
\item $x_{i}^{k}{\in}{\rm {\bf R}}(i{\in}S, k{\in}P, 0{\leq}x_{i}^{k}{\leq}1)$: a share of action $i$ in population $k$;
\item ${\Delta}=\left\{(z_{1},\ldots,z_{n})^{T}{\in}{\rm {\bf R}}^{n}\ {\mid}\ {\sum}_{j{\in}S}z_{j}=1,\ z_{i}{\geq}0\ {}^{\forall}i{\in}S\right\}$: the set of combinations of shares of actions in a population;
\item $x^{k}=(x_{1}^{k},\ldots,x_{n}^{k})^{T}{\in}{\Delta}$: a combination of shares of actions in population $k$;
\item $x=\left((x^{1})^{T},\ldots,(x^{m})^{T}\right)^{T}{\in}{{\Delta}^{m}}$: a combination of $x^{k}$;
\item $e_{i}{\in}{\Delta}(i{\in}S)$: the unit vector corresponding to action $i$;
\item $v^{k}(k{\in}P,0<v^{k}<1)$: a share of agents who belong to population $k$;
\item $y_{i}={\sum}_{k{\in}P}v^{k}x_{i}^{k}(0{\leq}y_{i}{\leq}1)$: a total share of action $i$ in all populations;
\item $y=(y_{1},\ldots,y_{n})^{T}{\in}{\Delta}$: a profile of the total share of each action in all populations;
\item int$({{\Delta}^{m}})$: the interior of ${{\Delta}^{m}}$;
\item $C(y)=\{i{\in}S\ {\mid}\ y_{i}>0\}$: the carrier of $y{\in}{\Delta}$;
\item $A^{k}{\in}{\rm {\bf R}}^{n{\times}n}$: the payoff matrix of the population $k$;
\item $a_{ij}^{k}$: the $ij$-th element of the payoff matrix $A^{k}$; and
\item $u^{k}(e_{i},e_{j})=e_{i}^{T}A^{k}e_{j}$: the payoff of agents with action $i{\in}S$ against $j{\in}S$.
\end{itemize}
\begin{figure}[t]
\begin{center}
\includegraphics[width=85mm,clip]{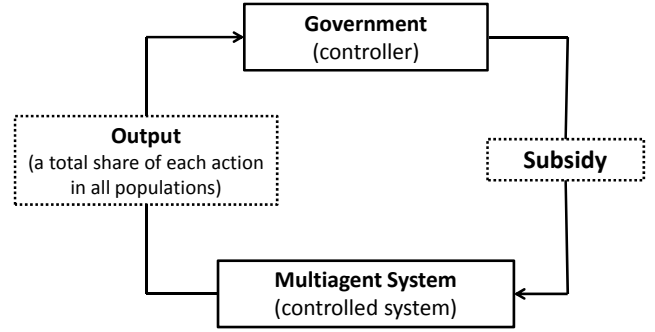}
\end{center}
\caption{Illustration of controlled multiagent system by government.}
\label{fig:two}
\end{figure}
We introduce a government that controls the multiagent system by observing a share of agents using each action not in each population, but in all populations. We call $x^{k}$ a population state of population $k$, and $y$ an output. Figure 3 shows the relationship between the population state combination $x$ and the corresponding output $y$. 
{\par}In Ref.\ [9], it is assumed that each agent's payoff depends on which population his opponent belongs to. However, in this paper, we assume that each agent's payoff does not depend on this. Since an agent plays a game with an opponent who selects action $i$ with probability $y_i$, the expected value of a payoff for an agent with action $i$ in population $k$ is 
\begin{equation*}
\underset{j{\in}S}{\sum}y_{j}u^{k}(e_{i},e_{j})=u^{k}(e_{i},y). 
\end{equation*}
The average payoff of all agents in the population $k$ is 
\begin{equation*}
\underset{i{\in}S}{\sum}x_{i}^{k}u^{k}(e_{i},y)=u^{k}(x^{k},y).
\end{equation*}
We assume that agents change their actions to obtain a higher payoff after the game. Also, we suppose that the increase rate of the share of action $i$ is proportional to the difference between the payoff which is earned with action $i$ and the average payoff in the population. Therefore, the replicator dynamics is given by the following differential equation:
\begin{equation}
\dot{x}_{i}^{k}=\left\{u^{k}(e_{i},y)-u^{k}(x^{k},y)\right\}x_{i}^{k},
\end{equation}
\noindent
for all $i{\in}S$ and $k{\in}P$. Equation (1) satisfies the following proposition:
\begin{proposition}
Equation (1) is invariant under any local shift of the payoff matrix $A^{k}$ for all $k{\in}P$, where the local shift is the addition of a constant to all elements of a column of payoff matrix $A^{k}.$ 
\end{proposition}

See Appendix A-1 for a proof of Proposition 1.
\begin{figure}[t]
\begin{center}
\includegraphics[width=85mm,clip]{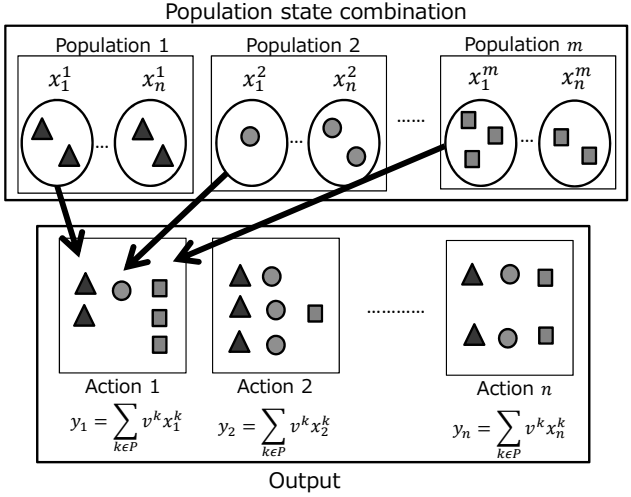}
\end{center}
\caption{The relationship between the population state combination $x$ and the corresponding output $y$.}
\label{fig:three}
\end{figure}

\section{The introduction of subsidy}
We introduce a government that controls the multiagent system by the subsidy. We assume that the government observes the output $y$ corresponding to the population state combination $x$. 
{\par}Let $y^{*}=(y_{1}^{*},\ldots,y_{n}^{*})^{T}$ be a target output of the government. Let $p_{i}$ be the number of agents that select action $i$, and $p={\sum}_{i{\in}S}p_{i}$ be the total number of agents of all populations. In this paper, the government determines the assignment of a subsidy to each agent based on the total amount of subsidies $D$, the target output $y^{*}$, and the current output $y$. We assume that, if $y_{i}^{*}=0$, then the subsidy to a group of agents with action $i$ is 0. If $y_{i}^{*}>0$, the government offers subsidy $Dy^{*}_{i}$ to a group of agents with action $i$, and equally distributes the subsidy among all agents in the group. Let $d>0$ be an average subsidy per agent. Note that $d$ is independent of the output $y$ and the population state combination $x$. Then, the total amount of subsidies is $D=dp$. We assume that $p_{i}>0$ for all $i{\in}S$. Thus, an agent with action $i$ obtains the subsidy $(Dy_{i}^{*})/p_{i}=(dpy_{i}^{*})/p_{i}=(dy_{i}^{*})/y_{i}$. After receiving the subsidy, the agent with action $i$ in population $k$ obtains the following payoff:
\begin{equation*}
\tilde{u}^{k}(e_{i},y)=u^{k}(e_{i},y)+df_{i}(y).
\end{equation*}
\noindent
The function $f_{i}:\bar{\Delta}{\rightarrow}{\rm {\bf R}}$ is defined as follows:
\begin{displaymath}
f_{i}(y) = \begin{cases}
\frac{y^{*}_{i}}{y_{i}} & \left(y_{i}^{*}>0\right), \\
0 & \left(y_{i}^{*}=0\right),
\end{cases}
\end{displaymath}
where
\begin{displaymath}
\bar{\Delta} = \left\{y{\in}{\Delta}\ {\mid}\ y^{*}_{i}>0\ {\Rightarrow}\ y_{i}>0\right\}.
\end{displaymath}
The average payoff in population $k$ is given by
\begin{eqnarray*}
\sum_{i{\in}S}x^k_i{\tilde{u}^k(e_i, y)} &=& \sum_{i{\in}S}x_{i}^{k}\left({u^{k}(e_{i},y)+df_{i}(y)}\right)\\
                       &=& u^{k}(x^{k},y)+d\left(\sum_{i{\in}S}f_{i}(y)x_{i}^{k}\right).
\end{eqnarray*}
\noindent
By substituting $\tilde{u}^{k}(e_{i},y)$ and $\sum_{i{\in}S}x^k_i{\tilde{u}^k(e_i, y)}$ for $u^{k}(e_{i},y)$ and $u^{k}(x^{k},y)$ in Eq.\ (1), respectively, the replicator dynamics with the subsidies is given by
\begin{equation}
\dot{x}_{i}^{k}=\left\{u^{k}(e_{i},y)-u^{k}(x^{k},y)\right\}x_{i}^{k}+dx_{i}^{k}\left(f_{i}(y)-\sum_{j{\in}S}f_{j}(y)x_{j}^{k}\right).
\end{equation}
\noindent
Equation (2) is invariant under any local shift of the payoff matrix $A^{k}$ for all $k{\in}P$, since the first term of Eq.\ (2) is the same as the right hand side of Eq.\ (1), and the second term is independent of the payoff function $u^{k}$. 
{\par}We suppose that $x(0){\in}{\rm int}({\Delta}^{m})$, where $x(0)$ is an initial population state combination. From the definition of $f_{i}(y)$, the domain of Eq.\ (2) is $\bar{\Delta}^{m}$, where
\begin{equation*}
\bar{\Delta}^{m}=\left\{x{\in}{\Delta}^{m}\ {\mid}\ y^{*}_{i}>0\ {\Rightarrow}\ {\sum}_{k{\in}P}v^{k}x^{k}_{i}>0\right\}.
\end{equation*}
{\par}We define $a_{\rm max}$, $a_{\rm min}$, $M_{i}$, and ${\Delta}^{m}_{\epsilon}$ as follows:
\begin{eqnarray*}
a_{\rm max} &=& \underset{i{\in}S, j{\in}S, k{\in}P}{\rm max}\left\{a_{ij}^{k}\right\},\\
a_{\rm min} &=& \underset{i{\in}S, j{\in}S, k{\in}P}{\rm min}\left\{a_{ij}^{k}\right\},\\
M_{i} &=& \frac{d}{a_{{\rm max}}-a_{{\rm min}}+d}y_{i}^{*}\ {\leq}\ y^{*}_{i},\\
{\Delta}^{m}_{\epsilon} &=& \left\{x{\in}{\Delta}^{m}\ {\mid}\ y^{*}_{i}>0\ {\Rightarrow}\ {\sum}_{k{\in}P}v^{k}x^{k}_{i}\ {\geq}\ {\epsilon}\right\}.
\end{eqnarray*}
Then, we have the following lemma:
\begin{lemma}
If $d>0$, then, for any $i{\in}C(y^{*})$, there exists $M_{i}>0$ such that $\dot{y}_{i}>0$ for all $y_{i}<M_{i}$.
\end{lemma}

See Appendix A-2 for a proof of Lemma1. From Lemma 1, we have the following proposition:
\begin{proposition}
If $0<{\epsilon}<{\rm min}_{i{\in}C(y^{*})}M_{i}$, then ${\Delta}^{m}_{\epsilon}$ is compact and invariant in Eq.\ (2).
\end{proposition}

See Appendix A-3 for a proof of Proposition 2. From Proposition 2, if $x(0){\in}{\rm int}({\Delta}^{m})$, then we can choose a sufficiently small ${\epsilon}<{\rm min}_{i{\in}C(y^{*})}M_{i}$, which satisfies $x(t){\in}{\Delta}^{m}_{\epsilon}$ for all $t{\in}[0,{\infty})$.
{\par}We define $\hat{X}$, $\bar{X}$, and $X^{*}$ as follows:
\begin{eqnarray*}
\hat{X}\ &=&\left\{x{\in}\bar{\Delta}^{m}\ {\mid}\ \left\{u^{k}(e_{i},y)-u^{k}(x^{k},y)\right\}x_{i}^{k}=0\ {}^{\forall}i{\in}S\ {}^{\forall}k{\in}P \right\},\\
\bar{X}\ &=&\left\{x{\in}\bar{\Delta}^{m}\ {\mid}\ {\sum}_{k{\in}P}v^{k}x_{i}^{k}=y_{i}^{*}\ {}^{\forall}i{\in}S \right\},\\
X^{*}&=&\hat{X}{\cap}\bar{X}.
\end{eqnarray*}
$\hat{X}$ is a set of equilibrium points of Eq.\ (1), and $\bar{X}$ is a set of population state combinations corresponding to $y^{*}$. If $y=y^{*}$ holds, then the second term of Eq.\ (2) is equal to zero. Therefore, the following proposition can be easily shown:
\begin{proposition}
$x^{*}{\in}X^{*}$ is an equilibrium point of Eq.\ (2) for all $d>0$.
\end{proposition}

From Proposition 3, the government chooses a target output $y^{*}$, and stabilizes the corresponding target equilibrium point $x^{*}{\in}X^{*}$ by offering the subsidies. In the following, we assume that there exists at least one element in $X^*$ which satisfies $y^{*}_{i}={\sum}_{k{\in}P}v^{k}x^{k*}_{i}$ for all $i{\in}S$.

\section{Global stabilization}


\subsection{Global stabilization condition}
We derive a sufficient condition for $x^{*}$ to be a globally asymptotically stable equilibrium point.
\begin{theorem}
We define two functions, $\bar{d}:\bar{\Delta}^{m}{\setminus}\bar{X}{\rightarrow}{\rm {\bf R}}$ and $F_{1}:\bar{\Delta}^{m}{\rightarrow}{\rm {\bf R}}$, as follows:
\begin{eqnarray*}
\bar{d}(x) &=& -\frac{\underset{k{\in}P}{\sum}v^{k}\left\{u^{k}(x^{k*},y)-u^{k}(x^{k},y)\right\}}{\underset{j{\in}S}{\sum}(y_{j}^{*}-y_{j})\frac{y_{j}^{*}}{y_{j}}},\\
F_{1}(x)       &=& {\sum}_{k{\in}P}v^{k}\left\{u^{k}\left(x^{k*},y\right)-u^{k}\left(x^{k},y\right)\right\}.
\end{eqnarray*}
If $X^{*}=\left\{x^{*}\right\},\ F_{1}(x)\ {\geq}\ 0$ for all $x{\in}\bar{X}$, and $d>{\rm max}\left\{0, {\rm sup}_{x{\in}\bar{\Delta}^{m}{\setminus}\bar{X}}\ \bar{d}(x)\right\}$, then $x^{*}$ is a globally asymptotically stable equilibrium point.
\end{theorem}
({\it Proof of Theorem 1})
{\par}Consider the following Lyapunov function candidate:
\begin{displaymath}
V(x)=\underset{k{\in}P}{\sum}\underset{i{\in}C(x^{k*})}{\sum}\left(-v^{k}x_{i}^{k*}{\rm log}\frac{x_{i}^{k}}{x_{i}^{k*}}\right).
\end{displaymath}
$V(x)>0$ holds for all $x{\in}\bar{\Delta}^{m}{\setminus}\{x^{*}\}$ and $V(x)=0$ holds if and only if $x=x^{*}$. The time derivative of $V(x)$ along solutions of Eq.\ (2) is
\begin{eqnarray*}
\dot{V}(x) &=& -\underset{k{\in}P}{\sum}\underset{i{\in}C(x^{k*})}{\sum}\frac{x^{k*}_{i}}{x^{k}_{i}}v^{k}\dot{x}^{k}_{i}\\
           &=& -F_{1}(x)-dF_{2}(x),
\end{eqnarray*}
where
\begin{eqnarray*}
F_{2}(x) &=& \underset{i{\in}C(y^{*})}{\sum}\left(y^{*}_{i}-y_{i}\right)\frac{y_{i}^{*}}{y_{i}}.
\end{eqnarray*}
We define $E$ as follows:
\begin{equation*}
E=\left\{x{\in}\bar{\Delta}^{m}\ {\mid}\ \dot{V}(x)=0\right\}.
\end{equation*}
We consider the case that $0<{\epsilon}<{\rm min}_{i{\in}C(y^{*})}M_{i}$ and $x{\in}\bar{\Delta}^{m}{\setminus}{\Delta}^{m}_{\epsilon}$. Then, from Lemma 1, $\dot{y}_{i}>0$ holds. Therefore, no equilibrium points exist in $\bar{\Delta}^{m}{\setminus}{\Delta}^{m}_{\epsilon}$. Moreover, if $y_{i}^{*}>0$ and $y_{i}<M_{i}$, then
\begin{eqnarray*}
\dot{x}^{k}_{i} &{\geq}& \left\{u^{k}(e_{i},y)-u^{k}(x^{k},y)\right\}x_{i}^{k}+dx_{i}^{k}\left\{\frac{y^{*}_{i}}{y_{i}}-1\right\}\\
                &>& \left(a_{\rm min}-a_{\rm max}\right)x^{k}_{i}+d\frac{x^{k}_{i}}{M_{i}}y_{i}^{*}-dx^{k}_{i}\\
                &=& \left(a_{\rm min}-a_{\rm max}\right)x^{k}_{i}+\left(a_{\rm max}-a_{\rm min}+d\right)x^{k}_{i}-dx^{k}_{i}\\
                &=&0.
\end{eqnarray*}
Therefore, we have $\dot{V}(x)<0$ for all $x{\in}\bar{\Delta}^{m}{\setminus}{\Delta}^{m}_{\epsilon}$, and $E$ is rewritten as follows:
\begin{equation*}
E=\left\{x{\in}{\Delta}^{m}_{\epsilon}\ {\mid}\ \dot{V}(x)=0\right\}=E_{1}{\cup}E_{2},
\end{equation*}
where
\begin{eqnarray*}
E_{1} &=& \left\{x{\in}\bar{X}\ {\mid}\ \dot{V}(x)=0\right\},\\
E_{2} &=& \left\{x{\in}{{\Delta}^{m}_{\epsilon}}{\setminus}\bar{X}\ {\mid}\ \dot{V}(x)=0\right\}.
\end{eqnarray*}
{\par}First, we consider the case that $x{\in}\bar{X}$. Obviously, we have ${\sum}_{k{\in}P}v^{k}x^{k}_{i}=y=y^{*}$, and from the assumption, $\dot{V}(x) = -F_{1}(x)\ {\leq}\ 0$ holds. Moreover, we have
\begin{eqnarray*}
E_{1} = \left\{x{\in}\bar{X}\ {\mid}\ \underset{k{\in}P}{\sum}v^{k}\left\{u^{k}\left(x^{k*},y^{*}\right)-u^{k}\left(x^{k},y^{*}\right)\right\}=0\right\}.
\end{eqnarray*}
The largest invariant set in $E_{1}$ is $X^{*}{\cap}E_{1}$ since the largest invariant set in $\bar{X}$ is $X^{*}$. We prove that $X^{*}{\cap}E_{1}=X^{*}$. From the definition of $\hat{X}$, for all $x^{*}{\in}\hat{X}$,
\begin{eqnarray*}
\left\{u^{k}(e_{i},y^{*})-u^{k}(x^{k*},y^{*})\right\}x^{k*}_{i}=0,
\end{eqnarray*}
holds for all $k{\in}P$ and all $i{\in}S$. Then, for all $k{\in}P$ and all $i{\in}C(x^{k*})$, $u^{k}(e_{i},y^{*})=u^{k}(x^{k*},y^{*})$ holds. Therefore, for all $x{\in}\bar{\Delta}^{m}$ and all $k{\in}P$, 
\begin{eqnarray}
C(x^{k}){\subset}C(x^{k*})\ {\Rightarrow}\ u^{k}(x^{k*},y^{*})-u^{k}(x^{k},y^{*})=0.
\end{eqnarray}
From the definition of $X^{*}$, for all $x{\in}X^*$, all $k{\in}P$, and all $i{\in}C(x^k)$, $u^{k}(e_{i},y^{*})=u^{k}(x^{k},y^{*})$ holds. Therefore, for all $x{\in}X^{*}$ and all $k{\in}P$,
\begin{eqnarray}
C(x^{k*}){\subset}C(x^{k})\ {\Rightarrow}\ u^{k}(x^{k*},y^{*})-u^{k}(x^{k},y^{*})=0.
\end{eqnarray}
From Eqs.\ (3) and (4), $x{\in}E_{1}$ holds for all $x{\in}X^{*}$. Therefore, for all $x{\in}\bar{X}$,
\begin{eqnarray}
X^{*}{\cap}E_{1}=X^{*}\ {\rm and}\ \dot{V}(x)\ {\leq}\ 0.
\end{eqnarray}
{\par}Next, we consider the case that $x{\in}{{\Delta}^{m}_{\epsilon}}{\setminus}\bar{X}$. Then, we have
\begin{eqnarray*}
F_{2}(x) &=& \underset{i{\in}C(y^{*})}{\sum}\frac{\left(y^{*}_{i}\right)^2}{y_{i}}-1.
\end{eqnarray*}
From Jensen's inequality,
\begin{eqnarray*}
-{\rm log}\left(\underset{i{\in}C(y^{*})}{\sum}\frac{\left(y^{*}_{i}\right)^2}{y_{i}}\right) &{\leq}& \underset{i{\in}C(y^{*})}{\sum}y^{*}_{i}\left(-{\rm log}\frac{y^{*}_{i}}{y_{i}}\right)\\
&{\leq}& {\rm log}\left(\underset{i{\in}C(y^{*})}{\sum}y^{*}_{i}\frac{y_{i}}{y^{*}_{i}}\right)\\
&{\leq}& {\rm log}\left(\underset{i{\in}S}{\sum}y_{i}\right)\\
&=&      {\rm log}(1)=0.
\end{eqnarray*}
Then, we have
\begin{eqnarray*}
\underset{i{\in}C(y^{*})}{\sum}\frac{\left(y^{*}_{i}\right)^2}{y_{i}}\ {\geq}\ 1,
\end{eqnarray*}
since ${\rm log}(y_{i})$ is a monotonically increasing function with respect to $y_{i}$. Thus, $F_{2}(x)=0$ holds if and only if $y=y^{*}$, and if $x{\in}{{\Delta}^{m}_{\epsilon}}{\setminus}\bar{X}$, then $F_{2}(x)>0$ holds. Therefore, we have
\begin{eqnarray*}
E_{2} &=& \left\{x{\in}{{\Delta}^{m}_{\epsilon}}{\setminus}\bar{X}\ {\mid}\ d=-\frac{F_{1}(x)}{F_{2}(x)}\right\}.
\end{eqnarray*}
Therefore, for all $x{\in}{{\Delta}^{m}_{\epsilon}}{\setminus}\bar{X}$,
\begin{eqnarray}
d>{\rm max}\left\{0, \underset{x'{\in}{\Delta}^{m}_{\epsilon}{\setminus}\bar{X}}{\rm sup}\ -\frac{F_{1}(x')}{F_{2}(x')}\right\}\ {\Rightarrow}\ E_{2}={\emptyset}\ {\rm and}\ \dot{V}(x)<0.
\end{eqnarray}
{\par}From Eqs.\ (5), (6), and LaSalle's invariance principle\ {\cite{key10}}, if $X^{*}=\left\{x^{*}\right\}$, $F_{1}(x)\ {\geq}\ 0$ for all $x{\in}\bar{X}$, and $d>{\rm max}\left\{0, {\rm sup}_{x{\in}{\Delta}^{m}_{\epsilon}{\setminus}\bar{X}}\ -\frac{F_{1}(x)}{F_{2}(x)}\right\}$, then $x^{*}$ is a globally asymptotically stable equilibrium point.\ \ \ \ \ \ \ \ \ \ \ \ \ \ \ \ \ \ \ \ \ \ \ \ \ \ \ \ \  \ \ \ \ \ \ \ \ \ \ \ \ \ \ \ \ \ \ $\Box$
{\par}In Ref. [10], we derived a global stabilization condition for $m=2$ and $n=2$ case. In this paper, we extend the condition to $m\ {\geq}\ 2$ and $n\ {\geq}\ 2$ case.
{\par}In the case that there exists more than one element in the set $X^*$, we cannot apply Theorem 1. So, it is future work to extend Theorem 1 to this case.

\subsection{Example}

We consider a 3-population and 2-action case. In this case, we can eliminate $x_{2}^{k}$ since $x_{1}^{k}+x_{2}^{k}=1$ for all $k{\in}P=\{1,2,3\}$. Thus, we select $x=\left(x^{1}_{1},x^{2}_{1}, x^{3}_{1}\right)^{T}$ as a state vector of the example. We suppose $x(0){\in}{\rm int}({\Delta}^m)$. We set $A^{1}$, $A^{2}$, $A^{3}$, $v^{1}, v^{2}$, and $v^{3}$ as follows:
\begin{eqnarray*}
A^{1}=
\begin{bmatrix}
2 & 1 \\
3 & 4
\end{bmatrix}
,\ 
A^{2}=
\begin{bmatrix}
3 & 1 \\
2 & 4
\end{bmatrix}
,\ 
A^{3}=
\begin{bmatrix}
3 & 4 \\
1 & 2
\end{bmatrix}
,\\ \\
v^{1}=0.2,\ v^{2}=0.3,\ {\rm and}\  v^{3}=0.5.
\end{eqnarray*}
In this case, from Eq.\ (1), if $d=0$, then $x=(0,0,1)^{T}$ and $x=(0,1,1)^{T}$ are asymptotically stable equilibrium points. Shown in Fig.\ 4 is a phase portrait for $d=0$. In this example, we set five initial state combinations: $x(0)=(0.01,0.01,0.01)^{T}$, $(0.01,0.99,0.01)^{T}$, $(0.99,0.01,0.01)^{T}$, $(0.99,0.99,0.01)^{T}$, and $(0.5,0.5,0.01)^{T}$. From Fig.\ 4, depending on the initial states, their trajectories converge to one of the stable equilibrium points.
\begin{figure}[t]
\begin{center}
\includegraphics[width=85mm,clip]{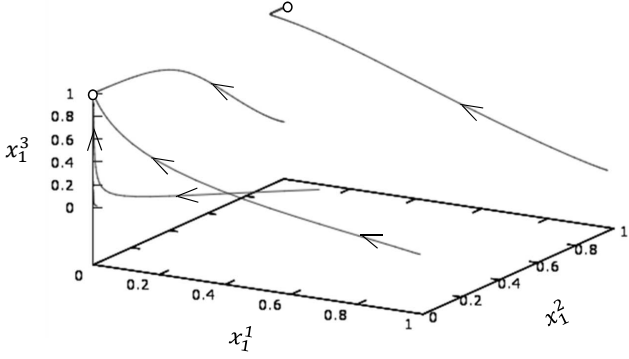}
\end{center}
\caption{Phase portrait for $d=0$.}
\label{fig:four}
\end{figure}
\begin{figure}[t]
\begin{center}
\includegraphics[width=85mm,clip]{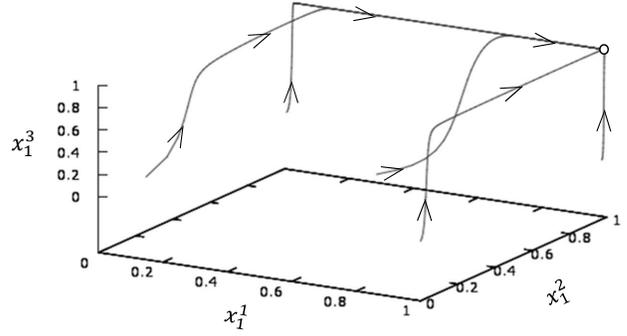}
\end{center}
\caption{Phase portrait for $d=1.2$ and $y^{*}=(1,0)^{T}$.}
\label{fig:five}
\end{figure}
{\par}First, we consider the case that $y^{*}=(1,0)^T$ is a target output on the boundary of the output space ${\Delta}$. Then, we have $X^{*}=\left\{(1,1,1)^T\right\}$ and the corresponding target equilibrium point is $x^{*}=(1,1,1)^T$. It is easily shown that ${\rm sup}_{x{\in}\bar{\Delta}^{m}{\setminus}\bar{X}}\bar{d}(x)<1.2$ and $F_{1}(x)>0$ for all $x{\in}\bar{X}$. Therefore, from Theorem 1, if $d=1.2$, then $x^{*}$ is a globally asymptotically stable equilibrium point.
{\par}Next, we consider the case that $y^{*}=(0.8, 0.2)^T$ is a target output in the interior of the output space ${\Delta}$. Then, we have $X^{*}=\left\{(0,1,1)^T\right\}$ and the corresponding target equilibrium point is $x^*=(0,1,1)^T$. It is easily shown that ${\rm sup}_{x{\in}\bar{\Delta}^{m}{\setminus}\bar{X}}\bar{d}(x)<1.5$ and $F_{1}(x)>0$ for all $x{\in}\bar{X}$. Therefore, from Theorem 1, if $d=1.5$, then $x^{*}$ is a globally asymptotically stable equilibrium point.
{\par}Shown in Figs.\ 5 and 6 are phase portraits for $y^{*}=(1,0)^{T}$ and $y^{*}=(0.8,0.2)^{T}$, respectively. The trajectories converge to $x^{*}=(1,1,1)^{T}$ and $x^{*}=(0,1,1)^{T}$ globally, respectively.
\section{Conclusion}

In this paper, we introduced a government that controls heterogeneous multiagent systems by using the subsidies, and described their controlled dynamics by the replicator dynamics. We also derived a global stabilization condition of a target equilibrium point in the $m$-population and $n$-action case $(m\ {\geq}\ 2, n\ {\geq}\ 2)$. 
{\par}Our future work is as follows:
\begin{itemize}
\item To derive a stabilization condition of a target output in the case that there exist several equilibrium points corresponding to the target output, or there exist no equilibrium points corresponding to the target output;
\item to discuss the robust stability of the target equilibrium point against uncertainties of the payoff matrix $A^k$;
\item to derive a stabilization condition of a target output in the case that the government does not know the population share $v^{k}$ or the payoff matrix $A^{k}$;
\item to discuss the stabilization of the target equilibrium point not by the subsidization but by the taxation; and
\item to apply our research to the control of social infrastructure such as smart grid.
\end{itemize}

\begin{figure}[t]
\begin{center}
\includegraphics[width=85mm,clip]{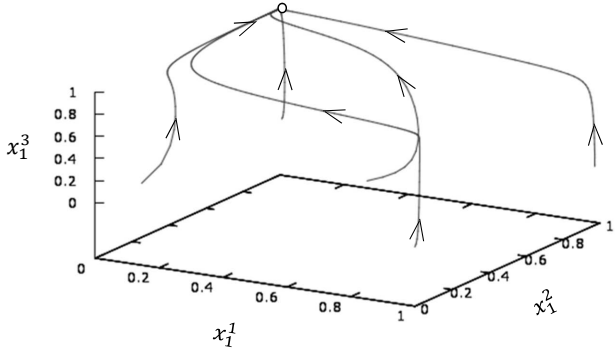}
\end{center}
\caption{Phase portrait for $d=1.5$ and $y^{*}=(0.8,0.2)^{T}$.}
\label{fig:six}
\end{figure}
\section*{Acknowledgment}
This work was supported in part by the Grant-in-Aid for Scientific Research (B) No.\ 24360164 and Young Scientists (B) No.\ 25820181.


\section*{Appendix A-1 : Proof of Proposition 1}
Let $\bar{A}^{k}$ be a payoff matrix where a constant $b$ is added to all elements of the column $j$ of $A^{k}$. Let $\bar{u}^{k}(s_{1},s_{2})=s_{1}^{T}\bar{A}^{k}s_{2}$. Then, $\bar{u}^{k}(e_{i},y)$ and $\bar{u}^{k}(x^{k},y)$ are given by
\begin{eqnarray*}
\bar{u}^{k}(e_{i},y) &=& e_{i}^{T}\bar{A}^{k}y \\
                     &=& a^{k}_{i1}y_{1}+{\ldots}+(a^{k}_{ij}+b)y_{j}+{\ldots}+a^{k}_{in}y_{n} \\ 
                     &=& u^{k}(e_{i},y)+by_{j},    \\ 
\bar{u}^{k}(x^{k},y) &=& ({x^{k}})^{T}\bar{A}^{k}y \\
                     &=& \underset{i{\in}S}{\sum}x_{i}^{k}e_{i}^{T}\bar{A}^{k}y \\ 
                     &=& \underset{i{\in}S}{\sum}x_{i}^{k}u^{k}(e_{i},y)+\underset{i{\in}S}{\sum}x_{i}^{k}by_{j} \\
                     &=& u^{k}(x^{k},y)+by_{j}.                        
\end{eqnarray*}
\noindent
Therefore, $\bar{u}^{k}(e_{i},y)-\bar{u}^{k}(x^{k},y)=u^{k}(e_{i},y)-u^{k}(x^{k},y)$ holds.\ \ \ $\Box$


\section*{Appendix A-2 : Proof of Lemma 1}
We define a function $g_{j}(w):{\rm {\bf R}}{\rightarrow}{\rm {\bf R}}$ as follows:
\begin{displaymath}
g_{j}(w)=\frac{y^{*}_{j}}{w}.
\end{displaymath}
Then, $f_{j}(y)=g_{j}\left(y_{j}\right)=g_{j}\left({\sum}_{k{\in}P}v^{k}x_{j}^{k}\right)$ holds. Thus, from Jensen's inequality,
\begin{displaymath}
g_{j}\left(\underset{k{\in}P}{\sum}v^{k}x_{j}^{k}\right)\ {\leq}\ \underset{k{\in}P}{\sum}v^{k}g_{j}\left(x_{j}^{k}\right).
\end{displaymath}
From Eq.\ (2), we have
{\small
\begin{eqnarray*}
\dot{x}_{i}^{k} &=& \{u^{k}(e_{i},y)-u^{k}(x^{k},y)\}x_{i}^{k}+dx_{i}^{k}\left\{\frac{y^{*}_{i}}{y_{i}}-\sum_{j{\in}S}g_{j}\left(\underset{k{\in}P}{\sum}v^{k}x_{j}^{k}\right)x_{j}^{k}\right\}\\
                &{\geq}& \{u^{k}(e_{i},y)-u^{k}(x^{k},y)\}x_{i}^{k}+dx_{i}^{k}\left\{\frac{y^{*}_{i}}{y_{i}}-\underset{j{\in}S}{\sum}\underset{k{\in}P}{\sum}v^{k}g_{j}\left(x_{j}^{k}\right)x_{j}^{k}\right\}\\
                &=& \{u^{k}(e_{i},y)-u^{k}(x^{k},y)\}x_{i}^{k}+dx_{i}^{k}\left\{\frac{y^{*}_{i}}{y_{i}}-1\right\}.
\end{eqnarray*}
}
Since $\dot{y}_{i}={\sum}_{k{\in}P}v^{k}\dot{x}_{i}^{k}$, we have
\begin{eqnarray*}
\dot{y}_{i} &{\geq}& \underset{k{\in}P}{\sum}v^{k}\{u^{k}(e_{i},y)-u^{k}(x^{k},y)\}x_{i}^{k}+dy^{*}_{i}-dy_{i}\\
            &{\geq}& \left(a_{{\rm min}}-a_{{\rm max}}\right)y_{i}+dy^{*}_{i}-dy_{i}\\
            &=&      \left(a_{{\rm min}}-a_{{\rm max}}-d\right)y_{i}+dy^{*}_{i}.
\end{eqnarray*}
Thus, if $y_{i}<\frac{d}{a_{{\rm max}}-a_{{\rm min}}+d}y_{i}^{*}=M_{i}$, then $\dot{y}_{i}>0$. Therefore, if $d>0$, then, for any $i{\in}C(y^{*})$, there exists $M_{i}>0$ such that $\dot{y}_{i}>0$ for all $x{\in}\bar{\Delta}^{m}$ with $y_{i}<M_{i}$.\ \ \ \ \ \ \ \ \ \ \ \ \ \ \ \ \ \ \ \ \ \ \ \ \ \ \ \ \ \ \ \ \ \ \ \ \ \ \ \ $\Box$


\section*{Appendix A-3 : Proof of Proposition 2}
Obviously, from the definition of ${\Delta}_{\epsilon}^{m}$, if $0<{\epsilon}<{\rm min}_{i{\in}C(y^{*})}M_{i}$, then ${\Delta}_{\epsilon}^{m}$ is compact. 
{\par}Next, we prove that if $0<{\epsilon}<{\rm min}_{i{\in}C(y^{*})}M_{i}$, then ${\Delta}_{\epsilon}^{m}$ is invariant in Eq.\ (2). Since ${\Delta}_{\epsilon}^{m}{\subset}{\Delta}^{m}$, if $x(0){\in}{\Delta}_{\epsilon}^{m}$, then ${\sum}_{i{\in}S}x_{i}^{k}(0)=1$ for all $k{\in}P$. From Eq.\ (2), ${\sum}_{i{\in}S}\dot{x}_{i}^{k}(t)=0$ for all $t{\in}[0,{\infty})$. Therefore,
\begin{equation}
{\sum}_{i{\in}S}x_{i}^{k}(t)=1\ \ {}^{\forall}k{\in}P\ \ {}^{\forall}t{\in}[0,{\infty}).
\end{equation}
From Eq.\ (2), we have
\begin{equation}
\left(\left(x_{i}^{k}=0\right){\vee}\left(x_{i}^{k}=1\right)\right)\ {\Rightarrow}\ \dot{x}_{i}^{k}=0.
\end{equation}
Moreover, from Lemma 1, if $y^{*}_{i}>0$ and ${\sum}_{k{\in}P}v^{k}x^{k}_{i}=y_{i}<M_{i}$, then $\dot{y}_{i}>0$. Therefore, we have
\begin{equation}
y_{i}(t)\ {\geq}\ {\epsilon}\ \ {}^{\forall}{\epsilon}<M_{i}\ \ {}^{\forall}t{\in}[0,{\infty}).
\end{equation}
From Eqs.\ (7), (8), and (9), if $0<{\epsilon}<{\rm min}_{i{\in}C(y^{*})}M_{i}$, then ${\Delta}_{\epsilon}^{m}$ is invariant in Eq.\ (2). \ \ \ \ \ \ \ \ \ \ \ \ \ \ \ \ \ \ \ \ \ \ \ \ \ \ \ \ \ \ \ \ \ \ \ \ \ \ \ \ \ \ \ \ \ \ \ \ \ $\Box$

%
\end{document}